\documentclass[12pt]{article}
\pdfoutput=1

\usepackage{paper2e}
\usepackage{mydefs2e}
\usepackage{xspace}
\usepackage{graphicx}

\newcommand{\Sla}[1]%
{\kern0.12em{\raise.15ex\hbox{$/$}\kern-.74em #1}}

\begin{document}

\begin{titlepage}

\title{Minimal Conformal Technicolor\\\medskip
and Precision Electroweak Tests}

\author{Jamison Galloway,\ \ 
Jared A. Evans,\ \
Markus A. Luty,\ \ 
and\ \ 
Ruggero Altair Tacchi}

\address{Physics Department, University of California Davis\\
Davis, California 95616}

\begin{abstract}
We study the
minimal model of conformal technicolor, an $SU(2)$
gauge theory near a strongly coupled conformal fixed point,
with conformal symmetry softly broken by technifermion mass terms.
Conformal symmetry breaking triggers chiral symmetry breaking
in the pattern $SU(4) \to Sp(4)$, which gives rise to a
pseudo-Nambu-Goldstone boson that can act as a composite
Higgs boson.
The top quark is elementary, and the top and electroweak gauge
loop contributions to the Higgs mass
are cut off entirely by Higgs compositeness.
In particular, the model requires no top partners and no
``little Higgs'' mechanism.
A nontrivial vacuum alignment results from the interplay of the
top loop and technifermion mass terms.
The composite Higgs mass is completely determined by the top loop, 
in the sense that $m_{h}/m_t$ is independent of the vacuum
alignment and is computable by a strong-coupling calculation.
There is an additional composite pseudoscalar $A$ with mass larger than
$m_{h}$ and suppressed direct production at LHC.
We discuss the electroweak fit in this model in detail.
Corrections to $Z \to \bar{b} b$ and  the $T$ parameter
from the top sector are suppressed by the enhanced
$Sp(4)$ custodial symmetry.
Even assuming that the strong contribution to the $S$ parameter is
positive and usuppressed,
a good electroweak fit can be obtained for $v/f \lsim 0.25$,
where $v$ and $f$ are the electroweak and chiral symmetry
breaking scales respectively.
This requires fine tuning at the $10\%$ level.
\end{abstract}

\end{titlepage}

\section{Introduction}
\label{sec:intro}
At the threshold of the LHC era, the origin of
electroweak symmetry breaking remains mysterious.
The two major experimental hints we have are the negative results
of searches for Higgs bosons,
and the constraints from precision electroweak measurements.
Known models that explain both of these facts
have either residual fine tuning or complicated structure
(or both).
Electroweak symmetry breaking by strong dynamics
is a theoretically compelling paradigm that naturally explains
the absence of a standard model Higgs boson,
but generally has difficulties explaining the precision electroweak data.
The other major problem with strong electroweak symmetry breaking
is explaining the quark and lepton masses without flavor-changing
neutral currents.

Conformal technicolor \cite{CTC} is a plausible paradigm
for addressing the flavor problem of strong
electroweak symmetry breaking.
The basic idea is that the strong sector is near a conformal fixed
point at high energy scales, with conformal invariance explicitly
but softly broken, for example by fermion mass terms.
Conformal symmetry breaking then triggers electroweak symmetry breaking.
This is a natural solution to the hierarchy problem in models where the
conformal breaking terms can be naturally small due to symmetries,
as for fermion mass terms.
This can help with the flavor problem because in a general conformal
fixed point the Higgs operator $\scr{H}$ whose VEV breaks
electroweak symmetry can have any dimension $d \ge 1$.
For $d = 1$ the operators that generate quark and lepton masses
scale like standard model Yukawa couplings, and flavor can be
decoupled to arbitrarily high scales.
However, for $d \to 1$ the correlation functions of $\scr{H}$ approach
those of a free field, and we recover the usual hierarchy problem.
This occurs because the dimension of $\scr{H}^\dagger \scr{H}$ approaches
$2d \to 2$ in this limit.
The idea is that $d = 1 + 1/\mbox{few}$, so that large anomalous
dimensions allow $\scr{H}^\dagger \scr{H}$ to be irrelevant,
while still allowing the flavor problem to be pushed to high scales.

In this paper, we consider the minimal model of conformal
technicolor, $SU(2)$ gauge theory with 2 electroweak doublets
with addition electroweak-singlet technifermions so that
the theory has a stongly-coupled conformal fixed point \cite{CTClattice}.
Mass terms for technifermions softly break the conformal symmetry
and give rise to electroweak symmetry breaking at low scales.
This model naturally has a composite Higgs boson,
and we will show that a good precision electroweak fit can be obtained
with very mild fine tuning.

The effective theory below the chiral symmetry breaking scale is determined by
the symmetry breaking pattern
$SU(4) \to Sp(4)$
(equivalent to $SO(6) \to SO(5)$),
which gives rise to 5 potential Nambu-Goldstone bosons.
Of these, 3 become the longitudinal components of the 
$W$ and $Z$ gauge boson.
The remaining 2 physical pseudo-Nambu-Goldstone bosons (PNGBs)
are a composite Higgs scalar $h$ and a pseudoscalar $A$.
The theory has a nontrivial vacuum alignment
parameterized by an angle $\theta$
defined by
\beq
v = f \sin \theta,
\eeq
where $v = 246\GeV$ is the electroweak vacuum expectation value (VEV)
and $f$ is the decay constant of the Nambu-Goldstone
bosons.
Here
$\sin\theta = 1$ corresponds to the ``technicolor'' vacuum
where electroweak symmetry is broken by the strong dynamics,
while for $\sin\theta \ll 1$ electroweak symmetry is broken
by the VEV of the composite Higgs.
The dominant contributions to the potential that determines $\theta$
naturally come from top loops and
from explicit technifermion masses.
The top loop completely determines the $h$ mass,
in the sense that
\beq
m_{h}^2 = c_t N_c m_t^2,
\eeq
where $N_c = 3$ is a color factor and $c_t \sim 1$ is an
effective coupling in the low-energy effective theory
that parameterizes how the Higgs compositeness cuts off the 
top loop.
The mass is independent of $\theta$, but it is
only for small $\theta$ that $h$ couples to electroweak gauge 
bosons like the standard model Higgs.
The other PNGB is the pseudoscalar $A$, which has mass
\beq
m_A^2 = \frac{m_{h}^2}{\sin^2\theta}.
\eeq
The pseudoscalar is therefore heavier than the scalar.
For $m_A > 2m_t$ the $A$ decays dominantly to $\bar{t}t$,
but for $150\GeV \lsim m_A < 2m_t$ it decays dominantly to
$WW$, $ZZ$, and $Z\ga$ (but not $\ga\ga$).
However, the $A$ couplings to electroweak gauge bosons and
tops are strongly suppressed compared to the analogous
Higgs couplings, and direct $A$ production is negligible
at the LHC.
It is possible that TeV resonances decaying to $A$ particles
may give an observable cross section,
but we leave the investigation of this to future work.

\begin{figure}[t]
\label{fig:STNDA}
\begin{center}
\includegraphics[scale=1]{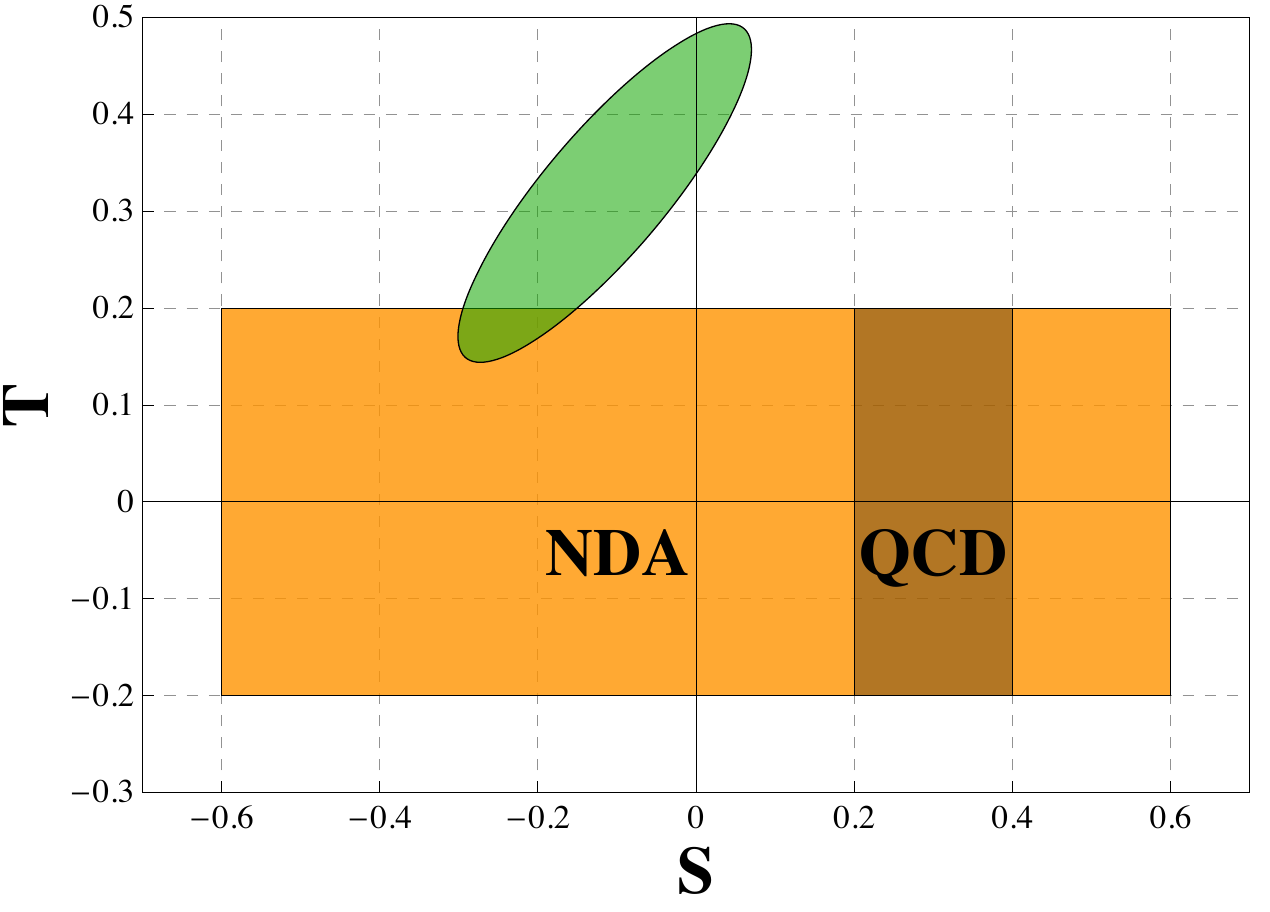}
\caption{ 
Expected range of $S$ and $T$ parameters in a general 
theory of electroweak symmetry breaking that is strongly
coupled at the TeV scale.
The reference Higgs mass is taken to be $1\TeV$.
The region denoted by NDA (``na\"\i{}ve dimensional analysis'')
is what is expected in a general theory of strong
electroweak symmetry breaking \cite{NDA}.
The region denoted by QCD is what is expected in a theory
of scaled-up QCD.}
\end{center}
\end{figure}

We investigate the precision electroweak
constraints on this model in some detail.
It is often stated that
models of strong electroweak symmetry breaking
are strongly disfavored by precision electroweak data.
The situation is illustrated in Fig.~\ref{fig:STNDA},
which shows the expectations for the $S$ and $T$ parameters
in a theory that is strongly coupled at the TeV scale,
with no light Higgs boson and no custodial symmetry
breaking other than the top mass and the gauging of $U(1)_Y$.
We also assume that the $S$ and $T$ parameters are not enhanced
by large $N$ factors.
The present model with $\sin\theta = 1$ satisfies all these
conditions. 
A good electroweak fit can be obtained with 
an additional positive contribution to $T$ and a small
or negative $S$ parameter.

Is it plausible that $S < 0$ in the present model?
Experimental data can be used to show that $S > 0$ in strong
electroweak sectors that are scaled-up
versions of QCD \cite{SQCD}.
However, it is noteworthy there is no theoretical argument that 
$S > 0$ for vectorlike gauge theories despite the fact that
many similar inequalities have been proven.
In gauge theories like QCD or the present theory, $S$ can be
expressed in terms of the physical states of the theory,
and is dominated by states with energies near the
strong coupling scale $\La$.
Conformal theories such as the one we are discussing are very
different from QCD above the scale $\La$, so there is no
reason to expect the states near the scale $\La$ to be
well-modeled by QCD.
It is therefore quite possible that $S < 0$ in this theory,
and the technicolor vacuum has a perfectly good electroweak fit.
The fact that strong electroweak symmetry breaking
may have a good electroweak fit
is much more general than the present model,
and should be taken seriously when assessing the plausibility
of strong electroweak symmetry breaking signals at the LHC.

Nonetheless, there are reasons to
suspect that $S > 0$ in strongly-coupled theories such as the
one under discussion.
In every known theory in which $S$ has a calculable sign, it 
turns out to be positive.
For example, simple extensions
of the standard model such as extra $SU(2)_W$ doublets give a
positive contribution to $S$.
More relevant for the present discussion are
5D ``Higgless'' models of electroweak symmetry breaking \cite{Higgless}.
These can be viewed as ``dual'' descriptions of large-$N$ strongly-coupled
electroweak symmetry breaking, and are conformal if the 5D spacetime
is AdS.
THese models also predict $S > 0$ when $S$ is calculable \cite{SHiggless}.

This motivates us to investigate whether there can be a good
precision electroweak fit if the contribution to $S$ from
the physics above the scale $\La$ is positive and unsuppressed
and unsuppressed,
\eg\ in the ``QCD'' region of Fig.~\ref{fig:STNDA}.
In the present model, away from the technicolor vacuum where $\sin\theta = 1$
there are additional negative contributions to the $S$ parameter from
loops of the composite Higgs boson.
These also give a positive contribution to the $T$ parameter,
which further improves the electroweak fit.
These contributions go in the right
direction because the theory becomes a (finely-tuned)
standard model for $\theta \ll 1$ with a light composite Higgs boson.
We will find that $\theta \lsim 0.25$ is sufficient to get a
good electroweak fit in the present model,
requiring fine tuning of order $\sin^2\theta \sim 10\%$.
It is also possible that there are additional contributions
to the $T$ parameter that allow even larger values of $\sin\theta$
(and less fine-tuning), but we confine ourselves to the minimal
model in this paper.

Composite Higgs models were first introduced in the context of
strongly-coupled gauge theories in the 1980's \cite{GKcompH}
and more recently revived in the context of 5D models 
\cite{compH5D}.
The present model is similar in spirit to the early composite Higgs
models, but it is based on a conformal rather than
an asymptotically free gauge theory.
The large coupling to the top quark is another important
new ingredient in the present model.
Asymptotically free $SU(2)$ gauge theories 
that give rise to the symmetry breaking pattern $SU(4) \to Sp(4)$
were considered as composite Higgs theories in the second paper
in \Ref{GKcompH}.
\Ref{intHiggs} analyzes a version of this theory where the top
quark is included and top partners are introduced to raise the scale
of compositeness above the TeV scale.
\Ref{GPS} analyzes a 5D model with the same coset, but considers
a different stabilizing potential with different phenomenology.
In the 5D models, the top loop contribution to the Higgs mass
are also off by top partners.
In the present model, the top quark contribution to the composite
Higgs mass is cut off entirely by compositeness of the Higgs sector,
and there is strong dynamics near the TeV scale.
The experimental signature of the top quark coupling
to the symmetry breaking sector is the presence of strong spin-0 resonances
coupling to the top quark \cite{ttbarTC}.

Another important difference between the present model 
and the 5D composite models is that the latter are effectively $1/N$ expansions
of a ``dual'' strongly-coupled large-$N$ theory.
In the 5D description, $N$ counts the number of KK modes below the
UV cutoff of the theory,
and hence parameterizes the range of validity of the 5D effective
theory.
These theories have a calculable positive contribution to the $S$
parameter proportional to $N$,
and this contribution must be canceled by fine-tuning.
This means that 5D composite Higgs theories are inevitably a
compromise between fine-tuning and predictivity of the effective theory.
This compromise is absent in the present theory, which has
a definite UV completion in terms of an $SU(2)$ gauge theory,
and has no large-$N$ enhancement of the $S$ parameter.
The UV completion is strongly-coupled, but may be investigated
on the lattice or using tools of conformal field theory.
Both are under active investigation and may give
nontrivial constraints on this class of models in the near future.

This paper is organized as follows.
In Section 2, we present the model
and review the constraints on the dimension of the
Higgs operator.
In Section 3, we determine the vacuum alignment, find
the spectrum and interactions of the PNGBs, 
and discuss the collider phenomenology
of the model.
In Section 4, we study the precision electroweak constraints
in this model.
Section 5 contains our conclusions.
Some technical results are collected in Appendices.

\section{The Model}
\label{sec:model}

\subsection{Minimal Conformal Technicolor}
We begin by defining the model \cite{CTClattice}.
It has a new strong $SU(2)$ gauge group,
with fermions transforming under
$SU(2)_{\rm CTC} \times SU(2)_W \times U(1)_Y$
as
\beq[TCsector]
\bal
\psi &\sim (2, 2)_0,
\\
\tilde{\psi}_1 &\sim (2, 1)_{-\frac 12},
\\
\tilde{\psi}_2 &\sim (2, 1)_{+\frac 12},
\\
\chi &\sim (2, 1)_0 \times 2n.
\eal\eeq
The fields $\psi$ and $\tilde\psi$ have the quantum numbers
of minimal technicolor \cite{minTC},
while $\chi$ are technifermions with no standard model charges.
There are $2n$ copies of $\chi$ fermions,
enough so that the theory has a strongly-coupled
conformal fixed point.
There is a growing body of lattice evidence for the existence
of such a fixed point in $SU(3)$ gauge theory near
12 flavors, although important disagreements between different groups remain
\cite{latticeCFT}.
There are currently no relevant
lattice simulations for $SU(2)$ gauge theory,
but comparison with supersymmetric theories leads one to
expect a strong conformal fixed point for $n \simeq 4$ \cite{SUSYfixed}.

The theory also contains 
mass terms for the technifermions:
\beq[tfermmass]
\De\scr{L} = 
-\ka \psi \psi -
\tilde{\ka} \tilde{\psi}_1 \tilde{\psi}_2 - 
K \chi\chi
+ \hc
\eeq
Although we refer to these as mass terms, the fermion bilinears
have a nontrivial scaling
dimension $1 < d < 3$, 
so $\ka$, $\tilde\ka$, and $K$ have
dimension $4-d$.
They are relevant perturbations, and they
take the theory out of the conformal fixed point at the scale
where they become strong.
We assume
$\ka, \tilde{\ka} \ll K$
so that conformal breaking is dominated by $K$.
The scale of confinement and conformal symmetry breaking is then
\beq
\La \sim K^{1/(4-d)}.
\eeq
At the scale $\La$, we assume that the theory is in the same
universality class as $SU(2)$ gauge theory with 4 fundamentals.
This can be motivated by extrapolating from a model where
the fixed point is weakly coupled, for example in the theory
with larger $n$ near the Banks-Zaks fixed point \cite{BZ}.
In this case, the theory is weakly coupled at the scale $K$,
and we can integrate out the massive technifermions perturbatively.
The effective theory below this scale is an $SU(2)$ gauge theory
with 4 fundamentals, which is asymptotically free and becomes strongly
coupled at a scale below $K$.
It is believed to confine and break chiral symmetry in analogy with QCD.
We assume that the picture is qualitatively similar in the strongly
coupled case, with the important difference that the coupling is
already strong at the scale $K$.

The theory has an approximate $SU(4)$ symmetry under which
the light fermions rotate into each other.
We therefore define a 4-component fermion vector
\beq[Psibasis]
\Psi = \pmatrix{\psi \cr \tilde{\psi}_1 \cr \tilde{\psi}_2 \cr}.
\eeq
The condensate is
\beq[thecondensate]
\avg{ \Psi^a \Psi^b} \propto \Phi^{ab} = -\Phi^{ba}
\eeq
with a constant of proportionality of order $\La^d$.
We assume that $\Phi$ has maximal rank, and has maximal symmetry,
in which case we have the symmetry breaking pattern
$SU(4) \to Sp(4)$.
This naturally generates electroweak symmetry breaking
via composite Higgs boson as we discuss below.

The scale $\La$ is put in ``by hand''
via the mass term $K$ in \Eq{tfermmass}.
This is an explicit but soft breaking of the
conformal symmetry, so it is natural for $K$ to be
small compared to fundamental scales such as the
Planck scale.
This gives a completely natural solution of the hierarchy
problem that is conceptually similar to softly broken
supersymmetry.
(In fact, both conformal symmetry and supersymmetry
are nontrivial extensions of Lorentz invariance,
so both involve a new broken spacetime symmetry.)
Soft supersymmetry
breaking may arise from spontaneous symmetry breaking in a more
fundamental theory,
explaining the origin of the soft breaking terms.
Similarly, the mass terms in \Eq{tfermmass} can arise from a
variety of different ways, including spontaneous breaking
of chiral symmetries, gauged or not.
We will not discuss the details here, and instead focus on the
phenomenology of the model at the TeV scale.

\subsection{Top Quark Mass}
The main phenomenological virtue of strong conformal dynamics above the TeV
scale is that it allows a plausible theory of
flavor, especially the origin of the top quark mass \cite{CTC}.
The top quark mass arises from a higher-dimension operator
of the form
\beq[topcoup]
\De \scr{L}_{\rm top} = \frac{g^2_t}{\La_t^{d-1}}
(Q t^c)^\dagger (\psi \tilde{\psi}_1) + \hc,
\eeq
where $d$ is the scaling dimension of 
$\psi\tilde\psi$ (the same as that of $\chi\chi$ since the theory
has a $SU(2n + 4)$ symmetry at the conformal fixed point)
and $g_t$ is a dimensionless coupling.
The low-energy physics depends only on the ratio $g_t^2 / \La_t^{d-1}$,
and it is convenient to choose $g_t$ so that
$\La_t$ is the scale where the operator gets strong.
The top mass is then given by
\beq
m_t \sim \La \left( \frac{\La}{\La_t} \right)^{d-1} \sin\theta,
\eeq
where the $\theta$ dependence arises because electroweak
symmetry is unbroken in the limit $\theta \to 0$.
Using
\beq
\La \sim 4\pi f \sim \frac{4\pi v}{\sin\theta},
\eeq
we have
\beq
\La_t \sim \frac{4\pi v}{\sin\theta}
\left( \frac{4\pi v}{m_t} \right)^{1/(d-1)}.
\eeq
Using $\sin\theta \sim 0.25$ and $4\pi v \sim 2\TeV$,
$m_t \simeq 165\GeV$ (the ``Yukawa mass''), we obtain
\beq
\La_t \sim \begin{cases}
30\TeV & $d = 3$, \cr
100\TeV & $d = 2$, \cr
1000\TeV & $d = 1.5$. \cr
\end{cases}
\eeq
For the free field theory value $d = 3$,
$\La_t \sim 3 \La$, and the
dynamics that generates the top mass cannot be disentangled
from that of electroweak symmetry breaking.
This possibility is explored in ``topcolor'' models \cite{topcolor}.
However, for $d < 3$ there can be a large separation of scales.
This is the basic idea of ``walking'' technicolor,
which goes back to the 1980's \cite{walking}.

\subsection{Bounds on Dimensions}
How small must $d$ be in order to have a sensible theory
of flavor?
The answer depends on what flavor symmetries (if any)
are preserved by the flavor dynamics.
The minimum requirement is that flavor dynamics occurs at a scale
at or below $\La_t$, the scale where the top quark coupling 
\Eq{topcoup} gets strong.
If the flavor dynamics satisfies minimal flavor violation,
then the scale of flavor physics may be quite low,
and we do not require very small values of $d$.
This may happen if the couplings of the standard model fermions
to the strong dynamics arises from the exchange of a heavy
scalar doublet with vanishing VEV.
Such a scalar may be natural in theories where supersymmetry is
broken at high scales, as in ``bosonic technicolor'' \cite{bosonicTC}
(see also \cite{scalarTC}).
In such a theory, the only flavor structure comes from the scalar
couplings to standard model fermions, so minimal flavor violation
is automatic.

If the theory of flavor does not have natural flavor conservation,
but only Yukawa suppression of flavor violation,
then we expect effective flavor-violating operators of the form
\beq
\De\scr{L}_{\rm eff}
\sim \frac{y_d y_s}{\La_f^2}  (\bar{s} d)^2
+ \cdots
\eeq
The strongest bound on these operators comes from the
$CP$-violating part of $K^0$--$\bar{K}^0$ mixing,
which gives
\beq
\La_f \gsim 30\TeV.
\eeq
The top quark coupling gets strong near $30\TeV$ for $d = 3$
(the free field value), so a strongly coupled theory of flavor
in principle may not require conformal dynamics.
Realistic theories of flavor probably require weak coupling,
in which case the minimal requirement is $d < 3$.
We leave detailed discussion of flavor models to future
work, but our point is that we do not necessarily
need very small values of $d$ to get a plausible theory
of flavor.

We now discuss theoretical constraints on the value of $d$.
One important tool is lattice studies.
The value of $d$ in strongly coupled gauge theories
can be measured by a lattice calculation using finite-size scaling
techniques \cite{CTClattice}.
These calculations are now starting to become a reality \cite{latticed}.

General results from conformal field theory also restrict the
value of $d$.
Unitarity implies
$d \ge 1$, and in the limit $d \to 1$ the correlation functions
of the operator becomes those of a free scalar \cite{CFTunitaritybound}.
This means that sufficiently close to $d = 1$ the ``Higgs''
operator ${\cal H} = \psi\tilde{\psi}$ becomes a weakly-coupled
scalar field, and we
get back the hierarchy problem of the standard model.
This can be stated in conformal field theory language
as follows \cite{CTC}.
The operator product expansion of $\scr{H}$ with itself
contains operators
\beq
\scr{H}^a(x) \scr{H}^\dagger_b(0)
\sim \de^a_b \, {\bf 1}
 +  \de^a_b \, [\scr{H}^\dagger \scr{H}](0)
+  (\si_i)^a_b \, [\scr{H}^\dagger \si_i \scr{H}](0)
+ \cdots
\eeq
The ``Higgs mass term'' $[\scr{H}^\dagger \scr{H}]$
is a singlet under all symmetries,
and therefore cannot be forbidden from
appearing in the effective Lagrangian.
To have a stable fixed point and avoid the hierarchy problem,
we therefore require
that it is an irrelevant operator:
\beq
\De = \dim[\scr{H}^\dagger \scr{H}] > 4.
\eeq
In the weakly-coupled limit $d \to 1$, we have
$\De \to 2$, and we recover the usual hierarchy problem.
The $d \to 1$ limit was investigated in detail in \Ref{CFTbounds1}.
They found that as $d \to 1$
\beq
\De \le 2d + \scr{O}((d-1)^{1/2}),
\eeq
so the limit is approached rather slowly.
The authors also derived quantitative limits on the
quantity
\beq
\De_{\rm min} = \min\!\left\{
\dim[\scr{H}^\dagger \scr{H}],\ 
\dim[\scr{H}^\dagger \si_i \scr{H}] \right\}.
\eeq
The techniques used do not distinguish operators that differ
only by internal symmetries, so they are not able to bound
$\De$.
The limits on $\De_{\rm min}$ are quite restrictive
\cite{CFTbounds1,CFTbounds2}, but it is important to
keep in mind that these bounds do not apply to the quantity of
interest for these models.

\section{Vacuum Alignment}
\label{sec:vacuum}
We now discuss the vacuum structure and electroweak symmetry
breaking in this model.
In the basis \Eq{Psibasis} the 
$SU(2)_L \times SU(2)_R$ generators are
\beq
T = \pmatrix{t_L & 0 \cr 0 & -t_R^T \cr},
\eeq
where $SU(2)_L = SU(2)_W$ and
$Y = T_{3R}$.
The condensate \Eq{thecondensate} then breaks the $SU(4)$
global symmetry down to $Sp(4)$.
Electroweak breaking depends on the alignment of the vacuum.
For example, there are vacua
\beq
\Phi \propto \pmatrix{\ep & 0 \cr 0 & \pm \ep \cr},
\qquad
\ep = \pmatrix{0 & 1 \cr -1 & 0 \cr}
\eeq
where $SU(2)_L \times SU(2)_R$ is unbroken.
We refer to these as ``electroweak vacua.''
We also have a vacuum where
\beq
\Phi \propto \pmatrix{0 & 1_2 \cr -1_2 & 0 \cr}
\eeq
where electroweak symmetry is maximally broken by the 
strong dynamics.
We refer to this as the ``technicolor vacuum.''

We will be interested in vacua between these limits.
In Appendix A, it is shown that the most general 
condensate up to $SU(2)_W$ transformations is given by
either $\Phi$ or $i\Phi$, where
\beq
\Phi = \pmatrix{e^{i\al} \cos\theta \, \ep & \sin\theta \, 1_2 \cr
-\sin\theta \, 1_2 & -e^{-i\al} \cos\theta \, \ep},
\eeq
with $0 \le \theta \le \pi$.
Note that the phase of the condensate is meaningful
once we require that the constant of proportionality in
\Eq{thecondensate} is real.
The condensates $\Phi$ and $i\Phi$ are not related by any symmetry of
the $SU(2)$ gauge theory, and are therefore physically distinct.
(In particular, they are not associated with degenerate vacua and
domain walls.)
Our final results do not depend on which choice represents the physical
vacua, and we will use $\Phi$.

The coset $SU(4)/Sp(4)$ contains 5 generators, 3 of which
correspond to the longitudinal polarizations of the massive
$W$ and $Z$.
The remaining 2 generators are physical PNGBs, which are
here parameterized by $\theta$ and $\al$.
We will be interested in $CP$ conserving vacua
where $\al = 0$.

The electroweak gauge boson masses arise from the covariant
kinetic term for the PNGBs:
\beq[PNGBkineticeff]
\scr{L}_{\rm eff} = \sfrac 12 f^2 \tr(
\Om^{\perp\mu} \Om^\perp_\mu)
= \sfrac 18 g^2 f^2 \sin^2\theta \, W^{+\mu} W^-_\mu + \cdots.
\eeq
(The formalism and notation is described in the Appendix.)
From this we see that the electroweak breaking scale is
\beq
v = f \sin\theta.
\eeq
The value of $\theta$ is determined by interactions
that explicitly break $SU(4)$.
The largest effects are
top loops, electroweak gauge boson loops,
and the explicit technifermion mass terms $\ka$ and $\tilde\ka$
in \Eq{tfermmass}.
These generate a potential for the PNGBs, which can also
be viewed as a potential for $\theta$.
We now discuss these contributions in turn.

\subsection{Technifermion Mass}
The technifermion mass terms \Eq{tfermmass} break
$SU(4)$ and therefore contribute to the PNGB potential.
We are assuming that $K > \ka,\tilde{\ka}$ so that $K$ triggers
chiral symmetry breaking.
Note that if the masses $K$, $\ka$, and $\tilde\ka$ have a common
origin, then it is natural for $\ka, \tilde{\ka} \lsim K$.
It is therefore natural for $\ka$ and $\tilde\ka$
to be important perturbations at the scale $\La$.
Note that this relies on the conformal technicolor mechanism
for generating the scale $\La$.
If $\La$ were a dynamical scale arising from a gauge coupling
becoming strong, then there would be no reason for mass terms
like $\ka$ and $\tilde\ka$ to be important perturbations at the scale
$\La$.

To keep track of the $SU(4)$ symmetry,
we write the small technifermion mass terms as
\beq
\scr{L} = -\Psi^T \scr{K} \Psi + \hc,
\eeq
where (in the basis \Eq{Psibasis})
\beq
\scr{K} = \pmatrix{\ka \ep & 0 \cr 0 & \tilde{\ka} \ep \cr}.
\eeq
We then view $\scr{K}$ as a spurion transforming under
$SU(4)$ as $\scr{K} \mapsto U^* \scr{K} U^\dagger$.
Treating $\scr{K}$ as a perturbation, the leading term in 
the potential in the effective theory below the scale
$\La$ is
\beq[Vmass]
V_{\rm mass} = \sfrac 14 \hat{C}_\ka \tr(\xi^T \scr{K} \xi \Phi) + \hc
= -\hat{C}_\ka (\ka - \tilde{\ka}) \cos\theta
+ \scr{O}(h, A)
\eeq
where $\xi$ parameterizes the PNGB fields (see appendices) and
\beq[Cm]
\hat{C}_\ka \sim \frac{\La^d}{16\pi^2}.
\eeq
The estimate \Eq{Cm} can be understood as follows.
The scale of the strong dynamics is set by $K$, and we can
normalize the technifermion fields so that $K \sim \La^{4-d}$.
Therefore, in the limit $\ka, \tilde{\ka} \to \La^{4-d}$,
all the technifermion mass terms are strong at the scale $\La$,
and the theory has a single scale.
In this limit, all contributions to the vacuum energy are expected
to be of order $\La^4 / 16\pi^2$, the same as the vacuum
energy of a free particle with mass $\La$.
This fixes the coefficient \Eq{Cm}. 
This argument is equivalent to 
``\naive dimensional analysis'' \cite{NDA}.

In fact, the effective coupling
$\hat{C}_\ka$ in \Eq{Vmass} is related to the
effective coupling that determines the top mass,
since both arise from the VEV of the technifermion bilinear.
The relation is
\beq[mtCkappa]
m_t = \frac{g_t^2 \hat{C}_\ka}{4\La_t^{d-1}}.
\eeq
where $g_t$ is the coefficient defined in \Eq{topcoup}.

\subsection{Top Loop}
The top quark coupling \Eq{topcoup} also breaks $SU(4)$,
and top loops give an important contribution to the PNGB
potential.
To keep track of the $SU(4)$ symmetry, we write the top quark
coupling \Eq{topcoup} as
\beq[topcoupsu4]
\De\scr{L}_{\rm eff} = \frac{g_t^2}{\La_t^{d-1}}
(Q t^c)^\dagger_\al \Psi^T P^\al \Psi
\eeq
where $\al = 1, 2$ is a $SU(2)_W$ index and
\beq[topproject]
P^1 = \frac 12 \pmatrix{0 & 0 & 1 & 0 \cr
0 & 0 & 0 & 0 \cr
-1 & 0 & 0 & 0 \cr
0 & 0 & 0 & 0 \cr},
\qquad
P^2 = \frac 12 \pmatrix{0 & 0 & 0 & 0 \cr
0 & 0 & 1 & 0 \cr
0 & -1 & 0 & 0 \cr
0 & 0 & 0 & 0 \cr}.
\eeq
We view $P^\al$ as a spurion transforming under $U \in SU(4)$ as
$P^\al \mapsto U^* P^\al U^\dagger$.
The leading effective potential induced by top loops is then
\beq[Vtop]
V_{\rm top} = -\sfrac 12
C_t \left| \tr(P^\al \xi \Phi \xi^T) \right|^2
=
-\sfrac 12 C_t \sin^2\theta + \scr{O}(h, A),
\eeq
where
\beq[Ct]
C_t \sim  \frac{N_c \La^4}{16\pi^2}
\left( \frac{\La}{\La_t} \right)^{2(d-1)}.
\eeq
Here $N_c = 3$ is the number of QCD colors.
The estimate \Eq{Ct} can be understood by noting that
$V_{\rm top} \sim N_c \La^4 / (16\pi^2)$ when $\La_t \to \La$,
the limit where all interactions are strong at the scale $\La$.
The sign of $C_t$ depends on the physics at the scale $\La$
and is not calculable.
We assume that $C_t > 0$, which is the sign obtained by computing
$V_{\rm top}$ in the low-energy effective theory with a
momentum cutoff at the scale $\La$.
This is also the sign obtained in models where the top loop
is cut off in a calculable way by top partners \cite{littleHiggsreview}.
The top loop then prefers to break electroweak symmetry.

The top quark mass depends on $\avg\theta$, the value of  $\theta$
at the minimum of the potential.
In order to keep the physical value of the top quark mass fixed,
the coupling of the top to the strong sector must depend on
$\avg\theta$.
To work this out, note that the top mass is contained in the term
\beq
\De\scr{L}_{\rm eff}
= a_t \La \left( \frac{\La}{\La_t} \right)^{d-1}
(Q t^c)^\dagger_\al \tr(P^\al \xi \Phi \xi^T) + \hc
\eeq
with $a_t \sim 1$ (so that $m_t \sim \La$ in the limit $\La_t \to \La$).
This gives
\beq
a_t \La \left( \frac{\La}{\La_t} \right)^{d-1}
\sin\avg\theta = m_t
\eeq
and hence
\beq
C_t \sim \frac{N_c \La^4}{16\pi^2}
\left( \frac{\La}{\La_t} \right)^{2(d-1)}
\sim \frac{N_c m_t^2 f^2}{\sin^2\avg\theta}.
\eeq
using $\La \sim 4\pi f$.

\subsection{Electroweak Gauge Loops}
The electroweak gauge couplings also break $SU(4)$ and
therefore contribute to the PNGB potential.
We view the gauge couplings and generators as spurions
transforming as $g_A T_A \mapsto U (g_A T_A) U^\dagger$,
where $A$ runs over the gauge bosons and generators.
The leading term in the effective potential is then
\beq
V_{\rm gauge} = C_g \sum_A g_A^2
\tr( \xi^\dagger T_A \xi \Phi \xi^\dagger T_A \xi \Phi)
= \sfrac 12 C_g (3 g^2 + g'{}^2) \sin^2 \theta
+ \scr{O}(h, A),
\eeq
where the sum is over the gauge generators and
\beq[Cgestimate]
C_g \sim \frac{\La^4}{(16\pi^2)^2}.
\eeq
This estimate can be understood from the fact that
$V_g \sim \La^4/16\pi^2$ in the strong coupling limit $g \to 4\pi$.
This contribution was estimated in \Refs{PeskinPreskill}
using a spectral representation for $C_g$ and assuming that
it is saturated by the lowest-lying vector resonances.
They obtained
\beq[CgPP]
C_g \simeq \frac{3 m_\rho^2 f^2}{16\pi^2}
\eeq
where $m_\rho$ is the mass of the lowest-lying vector
resonance.
This is consistent with the estimate \Eq{Cgestimate} since
$m_{\rho} \sim \La \sim 4\pi f$.
In particular $C_g > 0$, which results from vector meson dominance
and the assumption that the lowest-lying axial vector resonance 
has a mass larger than $m_\rho$ (as in QCD).
This favors the electroweak preserving vacuum.
However, we have seen that the top quark contribution has the same
$\theta$ dependence, and plausibly has the opposite sign.
These contributions appear to be comparable in size:
\beq
\frac{C_g (3 g^2 + g'{}^2)}{C_t} \sim
\frac{3 g^2 + g'^2}{N_c (m_t/v)^2}
\lsim 1.
\eeq
In models in which both the top and the gauge
loops are cut off in a calculable way, one finds that the top contribution
to the Higgs potential dominates \cite{littleHiggsreview} and
favors the electroweak symmetry breaking vacuum.
We will assume that this is the case also in the present model.

\subsection{Four-Technifermion Interactions}
The dynamics that generates the top quark interaction \Eq{topcoup} 
is expected to also generate four-technifermion interactions 
such as
\beq[Higgsmassterms]
\De\scr{L}_{4\psi} \sim \frac{g_{4\psi}^2}{\La_t^{\De - 4}}
|\psi \tilde{\psi}_1|^2 + \cdots
\eeq
where $\De$ is the dimension of the 4-fermion operator
and $g_{4\psi}$ is a dimensionless coupling that parameterizes
the strength of the four-technifermion interaction at the scale
$\La_t$.
In order to have a consistent IR fixed point, we require
these terms to be irrelevant, \ie $\De > 4$.
Estimating the size of these effects compared to top quark
loops, we obtain
\beq
\frac{V_{4\psi}}{V_{\rm top}}
\sim \left( \frac{g_{4\psi}}{4\pi} \right)^2
\left( \frac{m_t}{\La} \right)^{(\De - 2d - 2)/(d-1)}.
\eeq
We see that the four-technifermion interactions is suppressed
for small $g_{4\psi}$, and is even exponentially suppressed
provided that $\De > 2d + 2$.
This is model-dependent, and we will assume that
the top loop dominates in order to work with a minimal model.

\subsection{Minimizing the Potential}
Collecting the results above,
we find that the potential for
$\theta$ takes the simple form
\beq
V = -C_\ka \cos\theta - \sfrac 12 C_t \sin^2\theta,
\eeq
where
\beq
C_\ka = \hat{C}_\ka (\ka - \tilde{\ka}).
\eeq
Because we assume $C_t > 0$, the top contribution prefers the
``technicolor'' vacuum at
$\sin\theta = 1$, but the technifermion mass contribution
has a tadpole there.
The vacuum is therefore generally between the technicolor
and electroweak preserving vacuum,
exactly what is required to get a composite Higgs boson.

Minimizing the potential, we find extrema at
$\theta = 0$ and
\beq[thetheta]
\cos\theta = \frac{C_\ka}{C_t}.
\eeq
The extremum \Eq{thetheta}
has lower energy than $\theta = 0$
as long as $|C_\ka|/C_t < 1$.
We choose the technifermion masses so that 
$C_\ka > 0$, so that
$0 < \theta < \frac{\pi}{2}$.

\subsection{PNGB Spectrum}
We work out the masses and couplings of the physical PNGBs using
a basis of generators where the PNGB fields have vanishing
VEVs.
The generators then depend on $\theta$ (see Appendix B).
The generators corresponding to the scalar and
pseudoscalar PNGBs are
\beq[hgenerator]
X_h &= \frac{1}{2\sqrt{2}} \pmatrix{0 & i\ep \cr
i\ep & 0 \cr},
\\
\eql{agenerator}
X_A&= \frac{1}{2\sqrt{2}} \pmatrix{ \cos\theta & -\sin\theta \, \ep \cr
\sin\theta\, \ep & -\cos\theta \cr},
\eeq
respectively.
The masses of the PNGBs can be determined by expanding
the potential given above.
We find
\beq
m_h^2 = c_t N_c m_t^2,
\\
m_A^2 = \frac{m_h^2}{\sin^2\theta}.
\eeq
where
\beq
c_t = \frac{C_t \sin^2\avg\theta}{N_c m_t^2 f^2} \sim 1.
\eeq
Because of the special form of the potential,
$m_h / m_t$ is determined
completely by the top loop, and is independent of $\avg\theta$.
This can be understood from the fact that
$\theta$ can be used to parameterize the Higgs field,
so we have
\beq
m_h^2 = \frac{1}{f^2} \Avg{\frac{\partial^2 V}{\partial \theta^2}}
= \frac{C_t}{f^2} \sin^2\avg\theta
= N_c c_t m_t^2.
\eeq
This means that the Higgs mass in this model can in principle
be determined by a strong-coupling calculation of $C_t$,
which determines the top loop contribution to the
vacuum energy.
The fact that $m_A\to m_h$ as $\sin\theta \to 1$ can be
understood from the fact that there is an enhanced $U(1)$
custodial symmetry in this limit,
under which $h$ and $A$ form a complex charged multiplet.

\subsection{PNGB Interactions}
We now disucss the PNGB interactions with standard model
particles.
The leading couplings of the standard model gauge bosons
with the PNGBs can be read off from the kinetic term \Eq{PNGBkineticeff}:
\beq
g_{WWh} &= g_{WWh}^{\rm (SM)} \cos\theta,
\\
g_{ZZh} &= g_{ZZh}^{\rm (SM)} \cos\theta,
\\
g_{WWhh} &= g_{WWhh}^{\rm (SM)} \cos 2\theta,
\\
g_{ZZhh} &= g_{ZZhh}^{\rm (SM)} \cos 2\theta,
\\
g_{WWAA} &= g^2 \sin^2\theta,
\\
g_{ZZAA} &= (g^2 + g'^2) \sin^2\theta.
\eeq
The 4-point couplings are normalized so that the interaction terms 
are $\frac 14 g_{ZZAA} A^2 Z^\mu Z_\mu$, {\it etc\/}.

The PNGB kinetic term does not give rise to couplings of the form $AVV$,
where $V = W, Z, \ga$.
This is a consequence of $CP$ invariance, under which 
$A$ is odd (see Appendix C).
In order to have a $CP$ invariant $AVV$ coupling we need coupling
involving $\ep_{\mu\nu\rho\si}$.
Terms in the effective Lagrangian of the form
$\ep^{\mu\nu\rho\si}
\tr(\nabla_\mu \Om^\perp_\nu \Om^\perp_\rho \Om^\perp_\si)$
do not contain a $AVV$ vertex, and we must apparently go to
higher order in the derivative expansion.
In fact, the leading coupling comes from the chiral anomaly.
This is exactly analogous to the story for the $\pi^0 \ga\ga$
coupling in QCD, and we can use the standard derivation 
based on the chiral anomaly to find these couplings.
Writing the couplings as
\beq
\De\scr{L}_{\rm eff} = \sfrac 14
g_{A V_1 V_2} A \ep^{\mu\nu\rho\si}
V_{1\mu\nu} V_{2\rho\si}
\eeq
we have
\beq
g_{AWW} &= \frac{g^2 \sin\theta \cos\theta}{16 \sqrt{2} \pi^2 v},
\\
g_{AZZ} &= \frac{(g^2 - g'^2) \sin\theta \cos\theta}{16 \sqrt{2} \pi^2 v},
\\
g_{AZ\ga} &= \frac{g g' \sin\theta \cos\theta}{16 \sqrt{2} \pi^2 v}.
\eeq
Interestingly, $g_{A\ga\ga}$ vanishes at this order.
This is due simply to the vanishing of the trace of charges,
so this may be viewed as a consequence of the $SU(4)$ symmetry.

The PNGB fields also have couplings to the top quark via
the operator in the effective theory that contains the top quark
mass:
\beq
\De\scr{L}_{\rm eff} = \frac{m_t}{2\sin\avg{\theta}} (Q t^c)^\dagger_\al
\tr(\Phi \xi^T P_\al \xi) + \hc
\eeq
There are similar couplings for the remaining standard model
fermions, and we find
\beq
g_{h\bar{f}f} = g^{\rm (SM)}_{h\bar{f}f} \cos\theta.
\eeq
There is no contribution to $g_{A\bar{f}f}$ from the couplings above,
again due to the $SU(4)$ symmetry.
We get a nonvanishing coupling from higher order terms involving
explicit $SU(4)$ breaking.
For example, from the technifermion mass terms we have
\beq
\De\scr{L}_{\rm eff} \sim
\frac{m_t/\sin\avg\theta}{\La^{4-d}}
(Q t^c)^\dagger_\al
\tr(\xi^T K \xi \Phi \xi^T P^\al \xi \Phi) + \hc
\eeq
This gives 
\beq
g_{A\bar{t}t} \sim 
\frac{m_t}{f} \frac{\ka + \tilde{\ka}}{\La^{4-d}}
\sim \frac{N_c m_t^3}{16\pi^2 v^3} 
r \sin\theta \cos\theta,
\eeq
where
\beq
r = \frac{\ka + \tilde{\ka}}{\ka - \tilde{\ka}},
\eeq
and we have used the vacuum condition \Eq{thetheta} in the last step.
More generally, there is a coupling to all standard model fermions given by
\beq
g_{A\bar{f}f} = C r \frac{3 m_t^2 m_f}{16\pi^2 v^3} 
\sin\theta \cos\theta,
\eeq
where the normalization uncertainty is absorbed into an
overall factor $C$ that is the same for all fermions.

\begin{figure}[t]
\begin{center}
\includegraphics[scale=0.95]{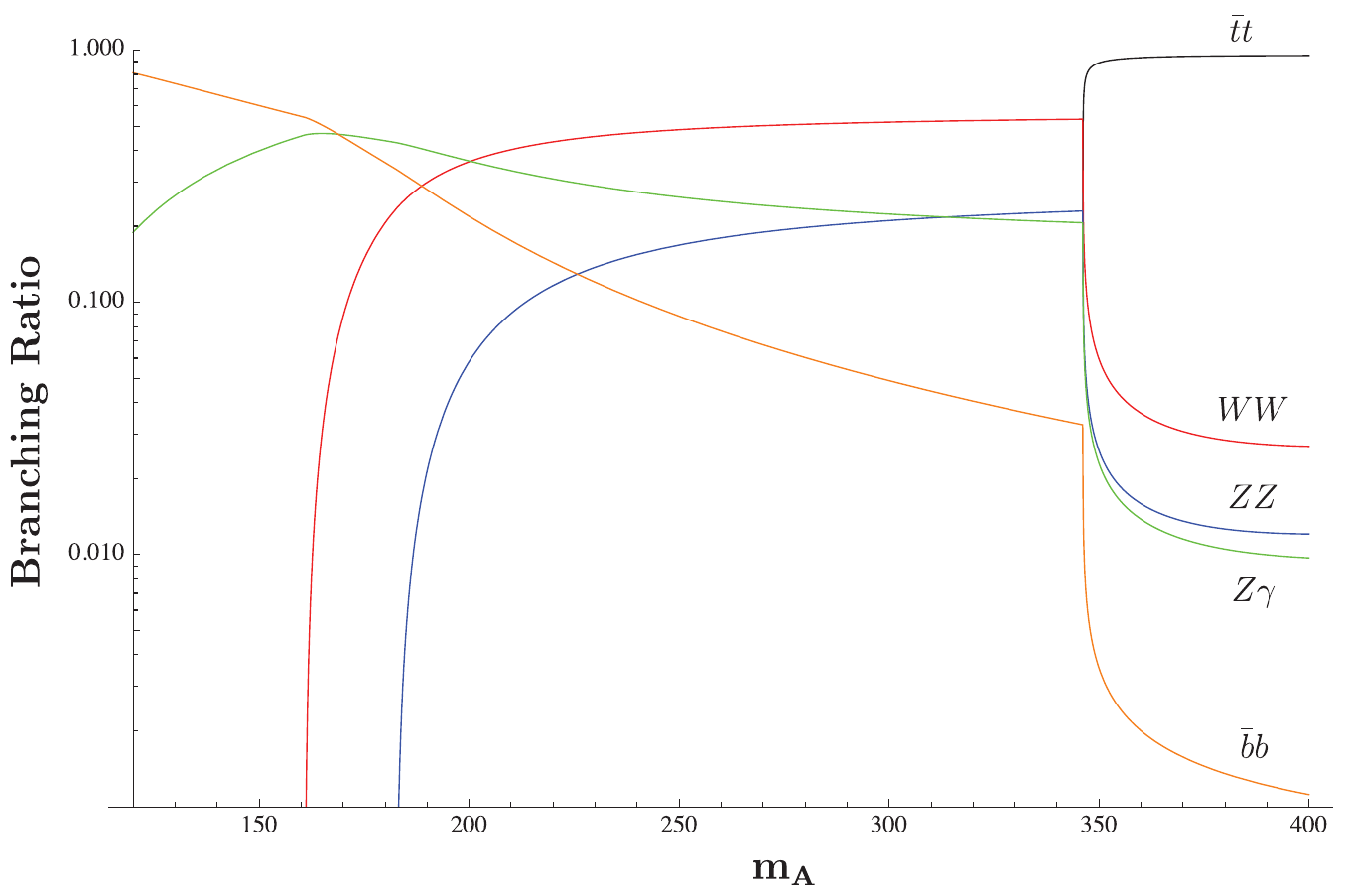}
\caption{\label{fig:ABR}
Branching ratios for $A$ decays
assuming $C r = 1$.
We take $m_h = 120\GeV$, so the value of $m_A$
fixes the value of the vacuum angle $\theta$.}
\end{center}
\end{figure}

We can use these couplings to compute the $A$ decays.
The gauge boson decay rates are given by
\beq
\Ga(A \to V_1 V_2) = \frac{g_{A V_1 V_2}^2}{32\pi}
\left[ m_A^2 - (m_{V_1} + m_{V_2})^2 \right]^{3/2},
\eeq
while the decay rate to fermions is
\beq
\Ga(A \to \bar{f}f) = \frac{N_c g_{A\bar{q}q}^2}{8\pi}
\left( m_A^2 - 4m_f^2 \right)^{1/2}.
\eeq
The branching ratios for $A$ decays are shown in Fig.~\ref{fig:ABR}.
It is important to remember that there is a large uncertainty in the
overall normalization of the couplings to fermions.
Nonetheless, it is clear that $A \to \bar{t}t$ dominates for
$m_A > 2m_t$, while decays to gauge bosons dominate for smaller
values of $m_A$ down to $m_A \sim 150\GeV$.

We now briefly discuss the PNGB phenomenology, leaving detailed
investigation for further work.
The phenomenology of the composite Higgs is similar to the
standard model, with suppressed couplings to the standard model particles.
This has been studied in detail in \Ref{SILH}.
The $A$ phenomenology is more distinctive.
Because the $A$ couplings to gauge bosons and top quarks are significantly
suppressed compared to the similar couplings of the standard model
Higgs, direct production of $A$ particles is tiny at the LHC.
However, the theory is expected to contain resonances at the scale
$\La \sim \mbox{TeV}/\sin\theta$, some of which may have decays involving
$h$ and $A$ particles.
These may give a significant production rate.
In particular, as shown in \Ref{ttbarTC}, we expect narrow spin-0
resonances at the scale $\La$ that can be efficiently produced via
gluon-gluon fusion through a top loop.
The phenomenology of strong resonances in this model
will be investigated in future work.

\section{Precision Electroweak Corrections}
\label{sec:PEW}
We now discuss the precision electroweak constraints coming
from the strong sector.
The potentially large contributions are
``oblique'' corrections from electroweak
gauge interactions, and a correction to
$Z \to \bar{b}b$ due to the coupling of the top with the
strong sector.

\subsection{$Z \rightarrow \bar{b} b$}
\begin{figure}[t]
\begin{center}
\includegraphics{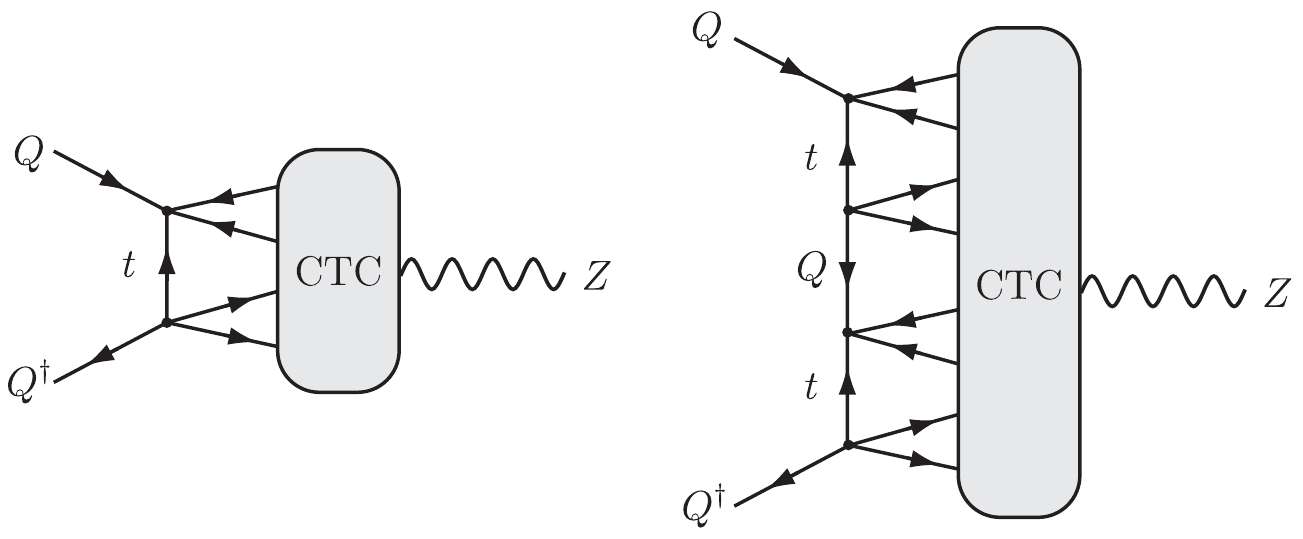}
\caption{\label{fig:Zbbar} 
Non-universal contributions
to $Z \to \bar{b}b$ from the top quark coupling
to the strong dynamics.}
\end{center}
\end{figure}

The coupling $g_{Z\bar{b}b}$ gets a non-universal contribution
due to the top mass operator \Eq{topcoup}.
There are contributions to $g_{Z\bar{b}b}$ from physics above the
scale $\La$ arise from the diagrams shown in Fig.~\ref{fig:Zbbar}.
The diagrams with two insertions of the top quark coupling
are parameterized in the low-energy effective theory by terms
of the form
\beq[Zbbar2]
\De\scr{L}_{\rm eff}
&\sim \left( \frac{\La}{\La_t} \right)^{2(d-1)}
Q^\dagger_\al \si^\mu Q^\be
\tr( \Om_\mu^\perp \xi^\dagger P^\dagger_\be 
P^\al \xi).
\eeq
\Eq{Zbbar2} gives a vanishing correction to $g_{Z\bar{b}b}$.
This can be traced to the $Sp(4)$ custodial symmetry.
Diagrams with four insertions of the top quark coupling give
rise to operators such as
\beq[Zbbar3]
\De\scr{L}_{\rm eff}
&\sim \left( \frac{\La}{\La_t} \right)^{4(d-1)}
Q^\dagger_\al \si^\mu Q^\be
\tr( \Om_\mu^\perp \xi^\dagger P^\dagger_\be
 P^\al P^\dagger_\ga P^\ga \xi),
\eeq
which give a tiny correction of order
\beq
\frac{\De g_{Z\bar{b}b}}{g_{Z\bar{b}b}} \sim
\left( \frac{m_t}{4\pi v} \right)^4 \sin^2 \theta
\sim 10^{-5} \sin^2\theta.
\eeq
PNGB loops do not contribute to $g_{Z\bar{b}b}$.
Contributions involving standard model gauge bosons are
included in the standard model contribution.
We conclude that non-universal corrections to
$Z \to \bar{b}b$ are small due to the  $Sp(4)$ custodial symmetry.

\subsection{$T$ Parameter}
The $T$ parameter measures the breaking of custodial $SU(2)$.
In this model, there is an enhanced $Sp(4)$ custodial symmetry
that further suppresses the $T$ parameter.
The $T$ parameter has potentially important
contributions from the top quark coupling,
the $U(1)_Y$ gauge interactions,
PNGB loops (including the composite Higgs),
and 4-technifermion
operators.
We will discuss each of these contributions in turn.

We begin with the top quark coupling.
The top quark contributes to the $T$ parameter via
a vacuum polarization involving a top quark loop.
This contribution is UV finite, meaning it is insensitive
to the physics at the scale $\La$.
This is included in the standard model fit to electroweak
data.
There are additional contributions to the $T$ parameter
from the top quark coupling \Eq{topcoup} arising from
physics above the scale $\La$.
In the effective theory below the scale $\La$, the terms arising
from two insertions of the top quark coupling that contribute to
the gauge boson masses have the form
\beq[topTop]
\De\scr{L}_{\rm eff} \sim \frac{\La^2}{16\pi^2}
\left( \frac{\La}{\La_t} \right)^{2(d-1)}
\Bigl[ &
\tr(\Om^{\perp \mu} \Om^\perp_\mu P^\dagger_\al P^\al)
\nonumber
\\
&+ \tr(\Om^{\perp \mu} \Om^\perp_\mu \Phi P^\al P^\dagger_\al \Phi^\dagger)
\\
&+ \tr(\Om^{\perp\mu} P^\dagger_\al (\Om^\perp_\mu)^T P^\al )
\Bigr].
\nonumber
\eeq
The first two terms do not contribute to the 
$T$ parameter because the custodial
$SU(2)$ breaking is $\De I = 1$.
The last term also gives a vanishing contribution to the $T$ parameter due
to the $Sp(4)$ custodial symmetry.
There are nonvanishing contributions to the $T$ parameter from
four insertions of the top quark interaction, \eg
\beq[topTopnonzero]
\De\scr{L}_{\rm eff} \sim \frac{\La^2}{16\pi^2}
\left( \frac{\La}{\La_t} \right)^{4(d-1)}
\tr(\Om^{\perp\mu} P^\dagger_\al P^\be \Om^\perp_\mu
P^\dagger_\be P^\al).
\eeq
The custodial symmetry violating mass term
$\De m_W^2 = m_{W3}^2 - m_{W^\pm}^2$
is proportional to $\sin^2\theta$, so we obtain
\beq
\frac{\De m_W^2}{m_W^2} \sim
\left( \frac{\La}{\La_t} \right)^{4(d-1)} \frac{g^2 f^2 \sin^2\theta}{m_W^2}
\sim \left( \frac{m_t}{4\pi v} \right)^4
\sim 10^{-5}.
\eeq
This gives a negligible correction $\De T = \al^{-1}
\De m_W^2 / m_W^2 \sim 10^{-3}$.

We now discuss the contribution to the $T$ parameter from
$U(1)_Y$ loops.
This arises from
\beq[TYeff]
\De\scr{L}_{\rm eff} \sim
\frac{g'^2 f^2}{16\pi^2}
\tr(\Om^{\perp\mu} \xi^\dagger Y \xi \Om^\perp_\mu 
\xi^\dagger Y \xi),
\eeq
where
\beq
Y = Y^\dagger = \pmatrix{0 & 0 \cr
0 & -\frac 12 \tau_3 \cr}
\eeq
is the $U(1)_Y$ generator.
The custodial symmetry violating mass term
is proportional to $\sin^4\theta$, and we obtain
\beq
\frac{\De m_W^2}{m_W^2} \sim
\frac{g'^2}{16\pi^2} \frac{ g^2 f^2 \sin^4\theta}{m_W^2}
\sim \frac{g'^2}{16\pi^2} \sin^2\theta
\sim 10^{-3} \sin^2\theta.
\eeq
We see that this contribution is suppressed at small $\sin\theta$.

There are also contributions from below the scale $\La$
due to PNGB loops.
The pseudoscalar does not contribute, while the Higgs
contributes
\beq\nonumber
\De T_{\rm IR} &= -\frac{3}{8\pi \cos^2\theta_W} \left[
\cos^2\theta \ln\frac{m_h}{\La}- \ln \frac{m_{h,{\rm ref}}}{\La} \right]
\\
\eql{TIR}
&= -\frac{3}{8\pi \cos^2\theta_W} \left[
\ln \frac{m_h}{m_{h,{\rm ref}}}
-\sin^2\theta \ln \frac{m_h}{\La} \right],
\eeq
where we have included the subtraction due to the reference
Higgs mass.
The first term in \Eq{TIR} is the usual standard
model contribution to the $T$ parameter from the Higgs loop.
The logarithmic dependence on $\La$ in the second term
in \Eq{TIR} represents a logarithmic UV divergence in the
effective theory below the scale $\La$.
The cutoff dependence is canceled by the $\La$
dependence of the effective couplings in
\Eq{TYeff}.

\subsection{$S$ Parameter}
The $S$ parameter gets corrections from all sources of electroweak
symmetry breaking.
The contribution from physics above the scale $\La$
are parameterized by the following 4-derivative terms in the
effective Lagrangian below $\La$:
\beq[Sct]
\De\scr{L}_S = 
-\sfrac 14 c_{\scr{F}} \tr( \scr{F}^{\mu\nu} \scr{F}_{\mu\nu} )
- \sfrac 12 c_{\scr{D}} \tr( \scr{D}^{\mu\nu} \scr{D}_{\mu\nu} )
\eeq
where
\beq
\scr{F}_{\mu\nu} = -i [\nabla_\mu, \nabla_\nu],
\\
\scr{D}_{\mu\nu} = \nabla_\mu \Om_\nu^\perp,
\eeq
where the covariant derivative is defined in Appendix C.%
\footnote{%
Because of the relations \Eq{Tpmeigenvalue} among the generators,
we do not get any new invariants by contracting $Sp(4)$ indices
using the $Sp(4)$ invariant metric $\Phi$.}
These contribute to the $S$ parameter
\beq[SUV]
S_{\rm UV} = 16\pi (\sfrac 12 c_{\scr F} - c_{\scr D}) \sin^2\theta.
\eeq
This represents the contribution to the $S$ parameter from
physics above the scale $\La$.

There are also contributions to the $S$ parameter from
physics below the scale $\La$, namely the PNGBs.
The pseudoscalar $A$ does not contribute to the $S$ parameter 
because it has only ``tadpole'' couplings to electroweak
gauge bosons, and therefore does not give a momentum-dependent
contribution.
The composite Higgs loop gives a contribution
\beq\nonumber
S_{\rm IR} &= \frac{1}{6\pi}
\left[ \cos^2\theta \ln\frac{m_h}{\La}- \ln \frac{m_{h,{\rm ref}}}{\La} \right]
\\
\eql{SIR}
&= \frac{1}{6\pi} \left[
\ln \frac{m_h}{m_{h,{\rm ref}}}
-\sin^2\theta \ln \frac{m_h}{4\pi v/\sin\theta} \right],
\eeq
where $m_{h,{\rm ref}}$ is the reference Higgs mass.
The first term in \Eq{SIR} is the usual standard
model contribution to the $S$ parameter from the Higgs loop,
and the logarithmic dependence on $\La$ in the second term
is canceled by the $\La$ dependence of the counterterms 
\Eq{Sct}.

\subsection{Electroweak Fit}
\begin{figure}[t]
\begin{center}
\includegraphics[scale=0.9]{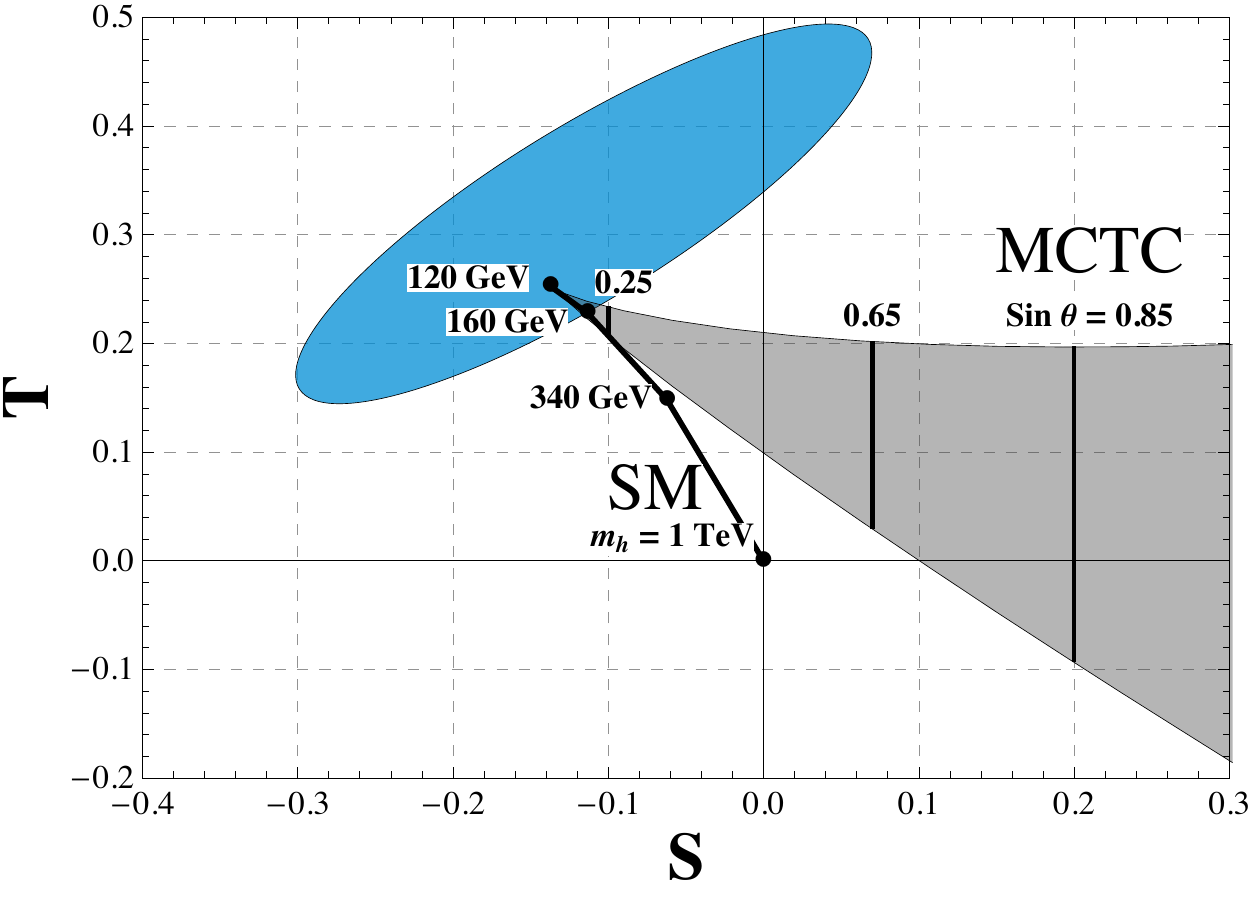}
\caption{\label{fig:STfit} 
Precision electroweak fit in the model described in the text
for $m_h = 120\GeV$.}
\end{center}
\end{figure}

We now put the results above together to discuss the fit to
precision electroweak data.
One important parameter is $m_h$, the mass of the scalar
PNGB.
Recall that $m_h$ is independent of $\theta$.
For $\sin\theta \ll 1$, the couplings of $h$ become those
of a composite Higgs boson,
and the precision electroweak fit reduces to the standard
model with Higgs mass $m_h$.
A good electroweak fit therefore requires $m_h^2 \lsim 180\GeV$.
Smaller values of $m_h^2$ allow larger values of $\sin\theta$
to get a good electroweak fit.
This is illustrated in Fig.~\ref{fig:STfit}.
This assumes $m_h = 120\GeV$, and that the UV contribution
to the $S$ parameter (see \Eq{SUV}) is positive and given by
the QCD value \cite{SQCD}, multiplied by $2/3$ to extrapolate
from $N_c = 3$ to $N_c = 2$.
We use the recent electroweak fit of \Ref{STrecent}.
Like the standard model, the present model has a single
parameter (in this case $\sin\theta$) that controls the precision
electroweak fit,
and has a good fit for a small range of this parameter.

However, the limit $\theta \ll 1$ is fine tuned,
and we must be close to this limit to get a good electroweak fit.
To quantify this tuning, we evaluate the sensitivity of the
electroweak VEV to the technifermion mass $\ka$,
a parameter in the fundamental theory that controls the
vacuum angle $\theta$.
We have
\beq
\mbox{sensitivity} = \frac{d \ln v^2}{d \ln \ka}
= -\frac{2}{\tan^2\theta}.
\eeq
As expected, this goes as $f^2/v^2 \sim \theta^{-2}$ for small
$\theta$.
For $\theta \sim 0.25$ the sensitivity is $\sim -30$.
The fine tuning is further reduced for smaller $m_h$.
Fine tuning may be completely absent if there are additional
positive contributions to the $T$ parameter.
In this case, we can allow $\sin\theta \lsim 0.5$,
which gives a sensitivity parameter $\sim 5$.

\section{Conclusions}
\label{sec:conclusions}
We have analyzed the minimal theory of conformal technicolor,
an $SU(2)$ gauge theory with fundamentals.
This gives a plausible theory of strong electroweak symmetry breaking,
including the possibility of decoupling flavor physics to high
scales.
We have analyzed the vacuum structure of this theory and showed
that it naturally gives rise to a vacuum with a composite Higgs
boson.
We analyzed the constraints from precision electroweak data and showed
that a good electroweak fit can be obtained with very mild fine tuning,
even if the strong contribution to the $S$ parameter is positive
and unsuppressed.
The minimal model of conformal technicolor may therefore sole the two
main problems of strong electroweak symmetry breaking.

The characteristic phenomenological features of this model
are a composite Higgs $h$
with mass close to the LEP bound,
and a heavier pseudoscalar $A$ decaying to $\bar{t}t$ or
$WW$, $ZZ$, and $Z\ga$ (but not $\ga\ga$).
The compositeness of the Higgs results in reduced $h$ couplings
to the standard model fields, as in all composite Higgs models.
Direct $A$ production is small at the LHC, but there may be
observable effects due to strongly-coupled heavy resonances
that decay to $A$ (and/or $h$).
In particular, the coupling of the top to the strong breaking
sector may allow may allow production of narrow spin-0 resonances
\cite{ttbarTC}.
The phenomenology of this model will be studied in detail in future
work.

Another area in which more work is needed is the flavor physics.
The flavor scale can be pushed to high scales if the dimension
$d$ of the technifermion bilinear is small, so the present
model can plausibly accommodate flavor.
However, it is important to have an explicit model for flavor
at high scales.
Such a model will show that
the flavor problem that has been
postponed to higher scales by large anomalous dimensions is
tractable.
An explicit flavor model is also necessary to
to check that flavor-changing neutral currents
are truly suppressed, and to determine the lowest allowed
value of the flavor scale.
This question will also be addressed in future work.

\section*{Acknowledgments}
We thank S. Chang, R. Rattazzi, and M. Schmaltz for 
healthy skepticism and stimulating discussions.
This work was partially carried out at the Aspen Center for Physics.
This work was supported by DOE grant
DE-FG02-91-ER40674.

\startappendices
\setcounter{section}{0}

\section*{Appendix A: The Space of Vacua}
\setcounter{section}{1}
In this appendix, we derive the most general form of the 
fermion condensate
\beq[condensate]
\avg{\Psi^a \Psi^b} \propto \Phi^{ab},
\eeq
where
\beq[Phiasymm]
\Phi^T = -\Phi.
\eeq
We define the phase of $\Phi$ by choosing the 
constant of proportionality 
in \Eq{condensate} to be real.

The space of vacua of this theory
is given by the space of possible VEVs of gauge
invariant operators such as \Eq{condensate}.
In particular, any VEVs related by symmetries
of the fundamental theory will correspond to states
with degenerate energy.
We assume that it is sufficient to study symmetry
transformations of the fermion condensate
to identify the space of vacua.
Global $SU(4)$ transformations act as
\beq[Phisu4trans]
\Phi \mapsto U \Phi U^T.
\eeq
Note in particular the transformation
\beq
\Psi \mapsto i\Psi,
\eeq
which maps $\Phi \mapsto -\Phi$.
This is a $Z_2$ transformation because the transformation
$\Psi \mapsto -\Psi$ is a gauge transformation.
We also have $CP$ transformations, under which
\beq
\Phi \mapsto \Phi^\dagger.
\eeq

The Pfaffian
\beq
\Pf(\Phi) = \sfrac 14 \ep_{abcd} \Phi^{ab} \Phi^{cd}
\eeq
is left invariant by $SU(4)$ transformations,
while under $CP$ it transforms as $\Pf(\Phi) \mapsto \Pf(\Phi)^*$.
We assume that the theory has $CP$ preserving vacua,
which implies that $\Pf(\Phi)$ is real.
We can then choose the normalization
of the condensate so that
\beq[PhiPf]
\Pf(\Phi) = \pm 1.
\eeq
Note that there is no symmetry of the theory that can
change the Pfaffian, so there is no reason that states
with opposite sign for $\Pf(\Phi)$ are degenerate in
energy.
The natural assumption is therefore that the physical
vacua correspond to one or the other sign.
We will refer to the cases $\Pf(\Phi) = \pm 1$
as ``$+$'' and ``$-$'' vacua, respectively.

We now assume that the vacuum preserves $Sp(4) \in SU(4)$.
This is satisfied if we impose
\beq[Phiunitary]
\Phi^\dagger \Phi = 1.
\eeq
To see this, it is sufficient to note that this is satisfied by
canonical $Sp(4)$ condensates such as
\beq
\Phi_0 = \pmatrix{0 & 1_2 \cr -1_2 & 0 \cr},
\eeq
and that this condition is
invariant under $SU(4)$.
Indeed, the defining relations \Eqs{Phiasymm},
\eq{PhiPf}, and \eq{Phiunitary} are all invariant under $SU(4)$
and $CP$ transformations.
Solving these constraints will therefore give us the 
most general condensate $\Phi$.
The most general antisymmetric matrix can be written
\beq
\Phi = \pmatrix{a \ep & c \cr -c^T & b \ep \cr}.
\eeq
where $A$ and $b$ are complex and $c$ is a $2 \times 2$
complex matrix.
\Eq{Phiunitary} implies
\beq
\eql{Ucondfirst}
c c^\dagger + |a|^2 1_2 = c^\dagger c + |b|^2 1_2 &= 1_2,
\\
\eql{Ucondsecond}
a c^\dagger \ep + b^* \ep c^T &= 0.
\eeq
\Eq{Ucondfirst} implies
\beq
c = r u
\eeq
where $u$ is unitary and $r$ is a real number given by
\beq
r = 1 - |a|^2 = 1 - |b|^2.
\eeq
We can use a $SU(2)_W$ transformation to set $u = e^{i\ga}$,
where $\ga$ is real.
Then \Eq{Ucondsecond} implies 
\beq
a e^{i\ga} = -(b e^{i\ga})^*.
\eeq
We conclude that up to $SU(2)_W$ transformations the most
general condensate satisfying \Eqs{Phiasymm} and \eq{Phiunitary} is
\beq
\Phi = e^{i\ga} \pmatrix{ e^{i\al} \cos\theta \, \ep & 
\sin\theta \, 1_2 \cr
-\sin\theta \, 1_2 &
-e^{-i\al} \cos\theta \, \ep \cr}.
\eeq
This has $\Pf(\Phi) = -e^{2i\ga}$, so the most general vacuum
is either $\Phi$ or $i\Phi$, where
\beq[Phiappendix]
\Phi = \pmatrix{ e^{i\al} \cos\theta \, \ep & 
\sin\theta \, 1_2 \cr
-\sin\theta \, 1_2 &
-e^{-i\al} \cos\theta \, \ep \cr}.
\eeq
We can change the sign of the block off-diagonal entry
with a $SU(2)_W$ (or $U(1)_Y$) gauge transformation,
so the physically distinct vacua are labelled by $\theta$ in the range
\beq
0 \le \theta \le \pi.
\eeq
Note that the vacua $\theta = 0$ and $\theta = \pi$ are related by the 
$Z_2$ transformation $\Psi \mapsto i\Psi$.

\section*{Appendix B: Generators}
\setcounter{section}{2}
This appendix derives the $Sp(4)$ and 
$SU(4)/Sp(4)$ generators associated with a general
$Sp(4)$ metric \Eq{Phiappendix}.
We can write an arbitrary $SU(N)$ generator as
\beq
T = T_\parallel + T_\perp,
\eeq
where
\beq[Spproject]
T_\parallel &= \sfrac 12 (T - \Phi T^T \Phi^\dagger),
\\
T_\perp &= \sfrac 12 (T + \Phi T^T \Phi^\dagger).
\eeq
This is a projection in the sense that
$(T_\parallel)_\parallel = T_\parallel$,
$(T_\parallel)_\perp = 0$,
{\it etc\/}.
These generators satisfy
\beq[Tpmeigenvalue]
T_\parallel \Phi + \Phi T_\parallel^T &= 0,
\\
T_\perp \Phi - \Phi T_\perp^T &= 0.
\eeq
This means that the $T_\parallel$ are $Sp(N)$ generators.
We identify $T_\perp$ with the broken generators of $SU(N)/Sp(N)$.
These definitions imply
$[T_\parallel, T_\perp] = \sum T_\perp$,
\ie the broken generators form a linear representation of
$Sp(N)$.
Also, the broken and unbroken generators are orthogonal in the
sense that
$\tr(T_\parallel T_\perp) = 0$.

To work out the explicit form of the generators it is useful
to start in the electroweak vacuum
\beq[EWvacstart]
\Phi_0 = \pmatrix{\ep & 0 \cr 0 & -\ep \cr}
\eeq
and perform a $SU(4)$ rotation to the general vacuum:
\beq
\Phi = U_0 \Phi_0 U_0^T
= \pmatrix{\cos\theta\, \ep & \sin\theta\, 1_2 \cr
-\sin\theta \, 1_2 & -\cos\theta\, \ep \cr},
\eeq
where
\beq
U_0 = \pmatrix{\cos\frac{\theta}{2}\, 1_2 & \sin\frac{\theta}{2}\, \ep \cr
\sin\frac{\theta}{2}\, \ep & \cos\frac{\theta}{2} \, 1_2 \cr}
\in SU(4).
\eeq
We can work out the generators for the vacuum \Eq{EWvacstart}
and rotate them into the general basis.
In the basis \Eq{EWvacstart}
the generators are
\beq
\!\!\!\!\!\!\!
T_{\parallel,0} = \pmatrix{a_i \si_i & c_0 + i c_i \si_i \cr
c_0 - i c_i \si_i & b_i \si_i \cr},
\quad
T_{\perp,0} = \pmatrix{x 1_2 & i y_0 + y_i \si_i \cr
-i y_0 + y_i \si_i & -x 1_2 \cr},
\eeq
where $a_i$, $b_i$, $c_0$, $c_i$, $x$, $y_0$, and $y_i$ are real.
Performing the transformation
\beq[generatorrotation]
T = U_0 T_0 U_0^\dagger,
\eeq
we find bases in a general vacuum state given by
\beq
\!\!\!\!\!\!\!\!
\bal
T_\parallel^1 &= \begin{pmatrix} { 0 & 1 \cr 1 & 0}\end{pmatrix},\  
T_\parallel^2 = \begin{pmatrix} { 0 & i \sigma_1 \cr -i \sigma_1 & 0}\end{pmatrix},\ 
T_\parallel^3 = \begin{pmatrix} { s 1_2 & c \epsilon \cr -c \epsilon & -s 1_2}\end{pmatrix},
\\
T_\parallel^4 &= \begin{pmatrix} { 0 & i \sigma_3 \cr -i \sigma_3 & 0}\end{pmatrix},\ 
T_\parallel^5 = \begin{pmatrix} { \sigma_1 & 0 \cr 0 & -\sigma_1}\end{pmatrix},\ 
T_\parallel^6 = \begin{pmatrix} { c \sigma_1 & s \sigma_3 \cr s \sigma_3 & c \sigma_1}\end{pmatrix},
\\
&\qquad\quad T_\parallel^7 = \begin{pmatrix} { \sigma_3 & 0 \cr 0 & -\sigma_3}\end{pmatrix},\ T_\parallel^8 = \begin{pmatrix} { c \sigma_3 & -s \sigma_1  \cr -s \sigma_1 & c \sigma_3}\end{pmatrix},
\\
&\qquad\quad T_\parallel^9 = \begin{pmatrix} { c \sigma_2 & -i s 1_2 \cr i s 1_2 & -c \sigma_2 }\end{pmatrix},
\ 
T_\parallel^{10} = \begin{pmatrix} { \sigma_2 & 0 \cr 0 & \sigma_2}\end{pmatrix},
\eal
\eeq
and
\beq
\!\!\!\!\!\!\!\!
\bal
T_\perp^1 &= \begin{pmatrix} { s \sigma_1 & -c \sigma_3 \cr -c \sigma_3 & s \sigma_1}\end{pmatrix},\ 
T_\perp^2= \begin{pmatrix} { s\sigma_2 & i c 1_2 \cr -i c 1_2 & -s \sigma_2}\end{pmatrix}, \ 
T_\perp^3= \begin{pmatrix} { s \sigma_3 & c \sigma_1 \cr c \sigma_1 & s \sigma_3}\end{pmatrix}, \ 
\\
&\qquad\qquad\quad
T_\perp^4 = \begin{pmatrix} { 0 & \sigma_2 \cr \sigma_2 & 0}\end{pmatrix}, \ 
T_\perp^5= \begin{pmatrix} { c1_2 & -s \epsilon \cr s \epsilon & -c 1_2}\end{pmatrix},
\eal
\eeq
where $s = \sin \theta$, $c = \cos \theta$.
The transformation \Eq{generatorrotation} preserves trace orthogonality, 
and it is easy to see that these generators are orthogonal.
In the technicolor limit $\theta \to \frac{\pi}{2}$ we have
\beq
T_\perp^i \to
\pmatrix{\si_i & 0 \cr 0 & \si_i^T \cr},
\qquad
i = 1, 2, 3,
\eeq
and we recognize these as the generators corresponding to the
longitudinal components of the $W$ and $Z$.
The physical PNGBs therefore correspond to the generators
$T^{4,5}_\perp$.
Under the preserved $CP$ symmetries, the PNGB fields transform
as $\Pi \mapsto -\Pi^*$ (see \Eq{CPPNGBtrans} below)
so the PNGB corresponding to $T^4_\perp$ ($T^5_\perp$) is
even (odd).
We therefore identify $T^4_\perp$ ($T^5_\perp$) with the generator
corresponding to $h$ ($A$).

\section*{Appendix C: Nonlinear Realization}
\setcounter{section}{3}
In this appendix, we review some details of the nonlinear
realization of $SU(4)/Sp(4)$.
The Nambu-Goldstone bosons can be described as broken
symmetry excitations of the order parameter  $\Phi$:
\beq
\Phi \to \xi \Phi \xi^T,
\eeq
where 
\beq
\xi = e^{i\Pi},
\eeq
with $\Pi$ a linear combination of the broken generators $T_\perp$
defined above.
The condensate $\Phi$ is $Sp(4)$ invariant:
\beq
\Phi = V \Phi V^T
\eeq
for all $V \in Sp(4)$.
We therefore define the transformation under $U \in SU(4)$
\beq
\xi \Phi \xi^T \mapsto U \xi \Phi \xi^T U^T = \xi' \Phi \xi'^T,
\eeq
where \cite{CWZ}
\beq[xitrans]
\xi \mapsto \xi' = U \cdot \xi \cdot V^\dagger(U, \xi).
\eeq
Here $V(U, \xi) \in Sp(4)$ is defined by the condition that
$\Pi'$ is a linear combination of broken generators.
Note that $V$ depends on $x$ through $\xi$.

We now construct the $SU(4)$ invariant terms in the effective
Lagrangian, following the standard construction of
\Ref{CWZ}.
We define the semi-covariant derivative
\beq
D_\mu \xi = \partial_\mu \xi - i \scr{A}_\mu \xi,
\eeq
where $\scr{A}_\mu$ are gauge fields for $SU(4)$.
This notation is appropriate for weakly gauging
all of $SU(4)$,
although we will only gauge the $SU(2)_W \times U(1)_Y$
subgroup.
These fields do not transform covariantly under
$SU(4)$ gauge transformations, but the quantity
\beq
\Om_\mu = i\xi^{\dagger} D_\mu \xi
\eeq
transforms like a $Sp(4)$ gauge field:
\beq
\Om_\mu \mapsto V (\Om_\mu + i \partial_\mu) V^{\dagger}.
\eeq
We project $\Om_\mu$ onto fields parallel and perpendicular
to the unbroken $Sp(4)$ direction using \Eq{Spproject}:
\beq
\Om_\mu^\parallel &= 
\sfrac 12 (\Om_\mu - \Phi \Om_\mu^T \Phi^\dagger),
\\
\Om_\mu^\perp &=
\sfrac 12 (\Om_\mu + \Phi \Om_\mu^T \Phi^\dagger).
\eeq
These then transform as
\beq
\Om_\mu^\parallel &\mapsto V (\Om_\mu^\parallel + i \partial_\mu) V^{\dagger},
\\
\Om_\mu^\perp &\mapsto V \Om_\mu^\perp V^{\dagger}.
\eeq
We therefore define the $Sp(4)$ covariant derivative
\beq
\nabla_\mu = \partial_\mu - i\Om_\mu^\parallel
\eeq
for fields transforming under $Sp(4)$.
For example, we can define
\beq
\scr{D}_{\mu\nu} = \nabla_\mu \Om_\nu^\perp
= \partial_\mu \Om^\perp_\nu - i
[\Om_\mu^\parallel, \Om_\nu^\perp]
\mapsto V \scr{D}_{\mu\nu} V^\dagger.
\eeq

We now discuss $CP$ invariance.
Under $CP$ the condensate transforms as
$\Phi \mapsto \Phi^\dagger$.
The condensate \Eq{Phiappendix} leaves invariant the combination of
this transformation and the
discrete symmetry $\Psi \mapsto i\Psi$,
which maps $\Phi \mapsto -\Phi = \Phi^T$.
The preserved $CP$ transformation is therefore
\beq
CP: \Phi \mapsto \Phi^*.
\eeq
Under this symmetry we have
$\xi \mapsto \xi^*$, and therefore
\beq[CPPNGBtrans]
CP: \Pi(x) \mapsto -\Pi^*(x^P),
\eeq
where $x^P$ is the parity transformed spacetime point.
If we impose the standard $CP$ transformation on the gauge fields
\beq
CP: A_\mu(x) \mapsto -A_\mu^*(x^P)
\eeq
we have the $CP$ transformations
\beq
\Om_\mu^\perp(x) \mapsto -[\Om_\mu^\perp(x^P)]^*,
\qquad
\Om_\mu^\parallel(x) \mapsto -[\Om_\mu^\parallel(x^P)]^*.
\eeq

\end{document}